\documentclass[useAMS,usenatbib,usedcolumn]{mn2e}

\usepackage{graphicx} 
\usepackage{times}

\title[Flux-density spectral analysis for several pulsars]{Flux-density spectral analysis for several pulsars and two newly-identified gigahertz-peaked spectra}
\author[M. Dembska et al.]{M. Dembska$^{1,2}$\thanks{e-mail: marta.dembska@dlr.de}, 
J. Kijak$^{1}$, A. Jessner$^{3}$, W. Lewandowski$^{1}$, B. Bhattacharyya$^{4}$ and 
Y. Gupta$^{5}$\\
$^{1}$ Kepler Institute of Astronomy, University of Zielona G\'ora, Lubuska 2, 65-265 
Zielona G\'ora, Poland \\
$^{2}$ German Aerospace Center, Institute for Space Systems, Robert Hooke Str. 7, 28359 Bremen, Germany \\
$^{3}$ Max-Planck-Institut f\"ur Radioastronomie, Auf dem H\"ugel 69, D-53121 Bonn, Germany\\
$^{4}$ Inter-University Centre for Astronomy and Astrophysics, Pune 411007, India\\
$^{5}$ National Centre for Radio Astrophysics, Pune University Campus, Postbag 3, India 411007}

\begin{document}

\date{Accepted\ldots Received\ldots ; in original form\ldots}

%\pagerange{\pageref{firstpage}--\pageref{lastpage}} \pubyear{2012}

\maketitle

\label{firstpage}

\begin{abstract}
In this paper we present results from flux density measurements for 21 pulsars over
a wide frequency range, using the Giant Metrewave Radio Telescope (GMRT) and the 
Effelsberg telescope.  Our sample was a set of mostly newly discovered pulsars from 
the selection of candidates for gigahertz-peaked spectra (GPS) pulsars. Using the 
results of our observations along with previously published data, we identify two new 
GPS pulsars. One of them, PSR J1740+1000, with dispersion measure of 24 pc cm$^{-3}$, 
is the first GPS pulsar with such a low DM value.  We also selected several strong 
candidates for objects with high frequency turnover in their spectra which require 
further investigation. We also revisit our source selection criteria for future 
searches for GPS pulsars.
\end{abstract}


\section{Introduction}
Of the about 2000 known pulsars, only around 400 have measured flux density spectra. 
For most of these cases, the flux density spectra can be described using a power-law 
of the form  
\begin{equation}
S(\nu)=d\cdot \nu^{\xi},
\label{powerlaw}
\end{equation}
where $\nu$ denotes observing frequency, with negative spectral index $\xi$ of about 
$-$1.8;  or by a combination of two power laws
\begin{equation}
S(\nu)=\left\{\begin{array}{ll} c_1\cdot \nu^{\xi_1}: & \nu\leqslant\nu_{b}\\ c_2\cdot \nu^{\xi_2}: & \nu>\nu_{b} \end{array}\right.
\label{break}
\end{equation}
with spectral indices $\xi_1$ and $\xi_2$ with average values of $-$0.9 and $-$2.2 
respectively, with a break frequency $\nu_b$, which is typically around 1.5 GHz \citep
{maron2000}. The population of pulsars with such break in the spectra is about 10\% of 
the sample analysed by \citet{maron2000}, and it is usually assumed that both $\xi_1$ 
and $\xi_2$ are negative with $\xi_2$ being steeper than $\xi_1$.  For pulsars which 
can be observed at low frequencies (i.e. about 100-600 MHz), a low frequency turnover 
is observed in the spectra~\citep{sieber73, lorimer95}. In some cases, positive 
spectral indices have also been observed between 400 MHz and 1600 MHz~\citep{lorimer95}. 

\citet{kijak2007} showed the first direct evidence of a high frequency turnover in 
radio pulsar spectra using multi-frequency flux measurements for candidates chosen 
from the objects which showed a decrease of flux density at frequencies below 
1 GHz. The frequency at which such a spectrum shows the maximum flux is called the peak 
frequency $\nu_{\rmn{p}}$.  Later, \citet{kijak2011b} presented a new class of 
objects, called gigahertz-peaked spectra (GPS) pulsars. These sources exhibit a
turnover in their spectra at high frequencies with $\nu_\rmn{p}$ about 1 GHz or 
above. To analyse the shapes of gigahertz-peaked spectra we use a function 
\begin{equation}
S(\nu)=10^{ax^2+bx+c},\quad x\equiv\log_{10}\nu,
\label{turnover}
\end{equation}
previously proposed for similar studies of pulsar spectra with turnover at low 
frequencies~\citep{kuzmin2001}. 

GPS pulsars have been demonstrated to be relatively young objects, and have a high 
dispersion measures ($DM\ge150$ pc cm$^{-3}$)~\citep{kijak2011a,kijak2011b}. They 
usually adjoin interesting and often dense objects in their vicinity, e.g. HII regions,
compact Pulsar Wind Nebulae (PWNe). In addition, it seems that they are coincident 
with known but sometimes unidentified X-ray sources from 3rd EGRET Catalogue or HESS 
observations. This in turn suggests that the high frequency turnover owes to the 
environmental conditions around neutron stars and/or the physical properties of the
interstellar medium. Though GPS pulsars represent the smallest group of 
radio pulsar spectra types today, \citet{bates2013} estimate that the number of such 
sources may be as much as 10\% of the whole pulsar population. It is obvious that 
without a more extensive sample of these sources, it is not possible to constrain 
reliably any statistics of their properties. 

\setcounter{table}{0}
\begin{table*}
\resizebox{\hsize}{!}{
\begin{minipage}{180mm}
\caption{The weighted means of all flux density measurements and their uncertainties 
for our sample of 21 pulsars are presented. Results from observations conducted in 2010 
using the GMRT in phased array mode at 610 MHz are marked as $S_{610}$; those using 
the Effelsberg telescope in 2012 at three frequencies 2.6 GHz, 4.85 GHz and 8.35 GHz 
are denoted by $S_{2600}$, $S_{4850}$ and $S_{8350}$, respectively. The total number 
of observations at each frequency are given in parentheses; "$\mathrm{<}$" denotes an 
upper limit. Flux density measurements from observations in 2008 are given in the 
footnote. Data collected in 2008 and 2010 were partially published (see \citeauthor
{kijak2011c}, \citeyear{kijak2011c}). Where possible, the value of $\xi$, the power 
index fitted to the spectra (using our data in combination with previously 
published data for individual pulsars), along with the reduced $\chi^2$, is given in the 
last two columns. There are no errors in the fits for $\xi$ for degenerate cases where 
the number of data points equals the number of parameters.}
\centering
\begin{tabular}{@{}l D{.}{.}{-1} c D{.}{.}{-1} D{,}{\pm}{4.4} D{,}{\pm}{4.4} D{,}{\pm}{6.5} D{,}{\pm}{4.4} D{.}{.}{3.3} D{.}{.}{2.2}@{}}
\hline
 \multicolumn{1}{c}{Pulsar} &  \multicolumn{1}{c}{Period} &   \multicolumn{1}{c}{DM} &   \multicolumn{1}{c}{Age} & \multicolumn{1}{c}{$S_{610}$} &  \multicolumn{1}{c}{$S_{2600}$} &   \multicolumn{1}{c}{$S_{4850}$} &   \multicolumn{1}{c}{$S_{8350}$} &  \multicolumn{1}{c}{$\xi$} &  \multicolumn{1}{c}{$\chi^2$}\\
            & \multicolumn{1}{c}{(s)} &  \multicolumn{1}{c}{$\left(\frac{pc}{cm^3}\right)$} &  \multicolumn{1}{c}{(kyr)}&  \multicolumn{1}{c}{(mJy)} &  \multicolumn{1}{c}{(mJy)} &  \multicolumn{1}{c}{(mJy)} & \multicolumn{1}{c}{(mJy)} & &\\
\hline

J1705$-$3950  & 0.319 & 207 & 83.4  & 1.2,0.1 (3) & & & &  0.3 &  \multicolumn{1}{c}{--} \\

J1723$-$3659 & 0.203 &  254 &  401 & 2.9,0.2 (2) & & & &  -0.8 & \multicolumn{1}{c}{--}\\

J1739$-$3023 &  0.114 &  170 & 159 &  2.7,0.2  (2) & & & & -1.2 & \multicolumn{1}{c}{--}\\

J1740+1000 &  0.154 &  24 &  114 &  &  2.0,0.7 (2) & 1.6,0.1 (2) & 0.344,0.014 (2) &  &  \\

J1744$-$3130 &  1.07 &  193 &  796  & 1.7,0.3 (2) & & & & -1.1 & \multicolumn{1}{c}{--} \\

J1751$-$3323 &  0.548 &  297 &  984 & 1.7,0.3  (2) & & & & -0.32 &  \multicolumn{1}{c}{--} \\

J1755$-$2521 & 1.18 & 252  & 207 & \multicolumn{1}{c}{$<$1.0 (2)} & & & & & \\

B1811+40 (J1813+4013) & 0.931 & 42  & 5790 &  & 0.185,0.036 (1) & & &  \multicolumn{1}{D{,}{\pm}{4.4}}{-1.2,0.1}  & 0.9\\

J1812$-$2102 &  1.22 & 547 &  811 &  \multicolumn{1}{c}{$<$2.7 (2)} & 0.38,0.07 (3) & 0.117,0.024 (3) & & \multicolumn{1}{D{,}{\pm}{4.4}}{-2.0,0.1} & 0.08\\

J1834$-$0731 &  0. 513 & 295  &  140 & \multicolumn{1}{c}{$<$1.6 (3)} & 0.406,0.009 (3) & 0.216,0.064  (3) & & \multicolumn{1}{D{,}{\pm}{4.4}}{-1.41,0.14} &0.6\\

J1835$-$1020 &  0.302 &  114 & 810 & 3.6,1.0  (3) & & & & -0.8  & \multicolumn{1}{c}{--}\\

J1841$-$0345 &  0.204 &  194 &  55.9 & 3.5,2.0 (2) &  0.979,0.015 (2) & 0.568,0.009 (1) & & \multicolumn{1}{D{,}{\pm}{4.4}}{-0.85,0.04} & 1.4\\

J1842$-$0905 & 0.345 & 343 &  520 & 2.2,0.3 (2) & & & & -1.2  & \multicolumn{1}{c}{--} \\

J1852$-$0635$^{1}$  &  0.524 & 171 &  567 & 7.0,0.7 (3)&  7.0,0.2 (3) & 4.686,0.044 (3) & 1.83,0.14 (3)  &  & \\

J1857+0143 & 0.140 &  249 & 71 &  \multicolumn{1}{c}{$<$1.0 (1)}   & & & &  &  \\

J1901+0510 &  0.615 &  429 &  313 & 3.4,1.5 (2) & & & & -1.8 & \multicolumn{1}{c}{--}\\

B1903+07 (J1905+0709) &  0.648 &  245 &  2080 &  & 0.59,0.12 (2) & & & \multicolumn{1}{D{,}{\pm}{4.4}}{-1.57,0.14}  & 4.2\\

B1904+06 (J1906+0641) & 0.267 &  473 & 1980 &  & 1.234,0.042 (1) & & & \multicolumn{1}{D{,}{\pm}{4.4}}{-0.94,0.18}  & 4.5\\

J1905+0616$^{2}$ &  0.99 &  256 &  116 & 2.4,0.5 (1) & 0.27,0.01 (3) & 0.203,0.023 (3) & & \multicolumn{1}{D{,}{\pm}{4.4}}{-0.88,0.25} & 4.5\\

J1910+0728 & 0.325 &  284 &  621 & 2.4,0.5 (2)  & 0.647,0.034 & & & \multicolumn{1}{D{,}{\pm}{4.4}}{-0.70,0.25} & 3\\

B1916+14 (J1918+1444) &  1.18 &  27 & 88.1  & &2.131,0.019 (1)  & & & \multicolumn{1}{D{,}{\pm}{4.4}}{-0.21,0.24} & 0.3\\

\hline
\multicolumn{10}{l}{$^{1}$ $S_{1170}^{2008}$=9.0$\rm \pm$0.5 (2), $\quad^{2}$ $S_{610}^{2008}$=2.1$\rm \pm$0.6 (1),  $S_{1170}$=0.4$\rm \pm$0.1 (2)}\\
  &  &  &  &  &  &  &  &  &\\
\end{tabular}
\label{tab_flux}
\end{minipage}
}
\end{table*}

The evolution of the spectrum of PSR B1259$-$63 with orbital phase can be treated as a 
key factor to define physical mechanisms which can be potentially responsible for the 
GPS phenomenon. The object is in a unique binary system with a massive (10M$_{\odot}$), 
main-sequence, Be star LS2883 with a radius of 6R$_{\odot}$. PSR B1259$-$63 is a 
middle-aged pulsar (330 kyr) with a relatively short period of 48 ms. Its average DM 
is about 147 pc cm$^{-3}$. The system is an eccentric binary ($e=0.87$) with an orbital 
period of 3.4 yr and a projected semi-major axis, $a\sin{i}$, of 1300 light seconds (2.6 AU). 
\citet{kijak2011a} used published data \citep{johnston99,johnston2005,connors2002} to 
show that at various orbital phases of the binary, the radio spectrum of PSR B1259$-$63 
mimics that of pulsars with the GPS.  The authors also presented evidence for evolution of 
the spectrum of the pulsar due to its orbital motion, along with the peak frequency 
dependence on the orbital phase. \citet{kijak2011a} proposed two possible causes of the 
observed variation in the spectra: free-free absorption in the stellar wind and cyclotron 
resonance in the magnetic field associated with a disk of the Be star. This observed 
evolution of the PSR B1259$-$63 spectrum seems to be the most 
convincing proof for an environmental origin of the GPS phenomenon.

\citet{kijak2013} studied the radio spectra of two magnetars PSRs J1550$-$5418 and 
J1622$-$4950, which clearly show turnover at frequencies of a few GHz. Both these 
magnetars are associated with supernova remnants, and thus are surrounded by a hot, 
ionized gas which can be responsible for the free-free absorption of radio waves. The 
authors concluded that the GPS feature in radio spectra of magnetars can be caused by 
external factors, in the same way as it occurs in the environment of GPS pulsars. 

Recently \citet{einstein} presented the spectrum of a newly discovered pulsar PSR 
J2007+2722 which shows a high frequency turnover (see Tab.~5 in their paper). It is 
possible that PSR J2007+2722 is a young pulsar that was born with a relatively weak 
magnetic field and with a birth period longer than it is usually assumed. In this case, 
with its $DM=127$ pc cm$^{-3}$, the object seems to be similar to other GPS pulsars. 
The distance of 5.4 kpc, corresponding to the DM value, can be the reason why there has 
not been mention about a SNR or a PWN around PSR J2007+2722.

In this paper we present flux measurements for 21 pulsars, resulting from observations 
at both low and high frequencies. The sample was chosen from objects suspected to be 
GPS pulsars. The results of our observations allow us to indicate several new sources 
showing GPS properties, and also strong GPS candidates. We concluded with a revisit of  
the source selection criteria for future searches for GPS pulsars. 

\section{Observations and results}
We present in this paper the flux density measurements collected during two observing 
projects.  We were able to successfully measure the flux densities of 12 pulsars at 610 
MHz and to estimate upper limits for another 4 pulsars, from observations in May 2010 
using the Giant Metrewave Radio Telescope (GMRT, near Pune, India), see also~\citet
{kijak2011c}. We also used the 100-m Effelsberg Radio Telescope for flux density 
measurements of 11 pulsars at 2.6 GHz, 4.85 GHz and 8.35 GHz in November 2012. The
main purpose of these observations was to find more pulsars showing gigahertz-peaked 
spectra.

We chose the sources to be observed with the GMRT from a set of newly discovered pulsars 
with periods between 0.1 and 2.0 seconds, whose flux density measurements at 1.4 GHz are 
available in the ATNF catalogue\footnote{http://www.atnf.csiro.au/research/pulsar/}. The 
observations were conducted in the phased array mode of the GMRT~\citep{gupta2000}, with 
the available 16 MHz bandwidth divided into 256 channels, and using a sampling rate of 
0.512 ms. The typical integration time was about 30 min (for more details see \citeauthor
{kijak2011b} \citeyear{kijak2011b}).

For the Effelsberg observing project we selected pulsars based on our GMRT observations, 
along with data from the ATNF catalogue and~\citet{maron2000}. Objects which appeared 
to have a flat spectrum or a positive spectral index below a frequency of 1 GHz were 
selected. To ascertain the full shape of the spectrum it was necessary to measure the 
flux densities at frequencies above 2 GHz. For these observations, we used the secondary 
focus receivers (with cooled HEMT amplifiers) with bandwidth of 100 MHz at 2.6 GHz, 
500 MHz at 4.85 GHz and 1100 MHz at 8.35 GHz. These receivers provided circularly 
polarised LHC and RHC signals, digitized and independently sampled with 1024 bins per 
period, and synchronously folded using the topocentric pulsar rotational period~\citep
{jessner96}. The typical integration time was around 20-25 min. The observations were 
intensity calibrated by the use of a noise diode and known continuum radio sources 
(NGC 7027, 3C 286, 3C 345). 

Table~\ref{tab_flux} combines the weighted means of all flux density measurements and 
their uncertainties for our sample of 21 pulsars, acquired at different frequencies with 
the GMRT and the Effelsberg observing projects. Most sources in our sample have high DMs 
and hence the flux density variations caused by interstellar scintillations should be 
minimal. Nevertheless, observations at each frequency (with a few exceptions) were 
conducted at 2-3 different epochs to give us reliable estimates of the average flux and 
its uncertainties, which are used to construct the spectra for these pulsars, in 
combination with flux density data available from literature. In Tab.~\ref{tab_flux} 
we also present other basic parameters of these pulsars, such as age, dispersion 
measure and period.

\begin{figure}
\begin{flushleft}
\includegraphics[width=8.2cm]{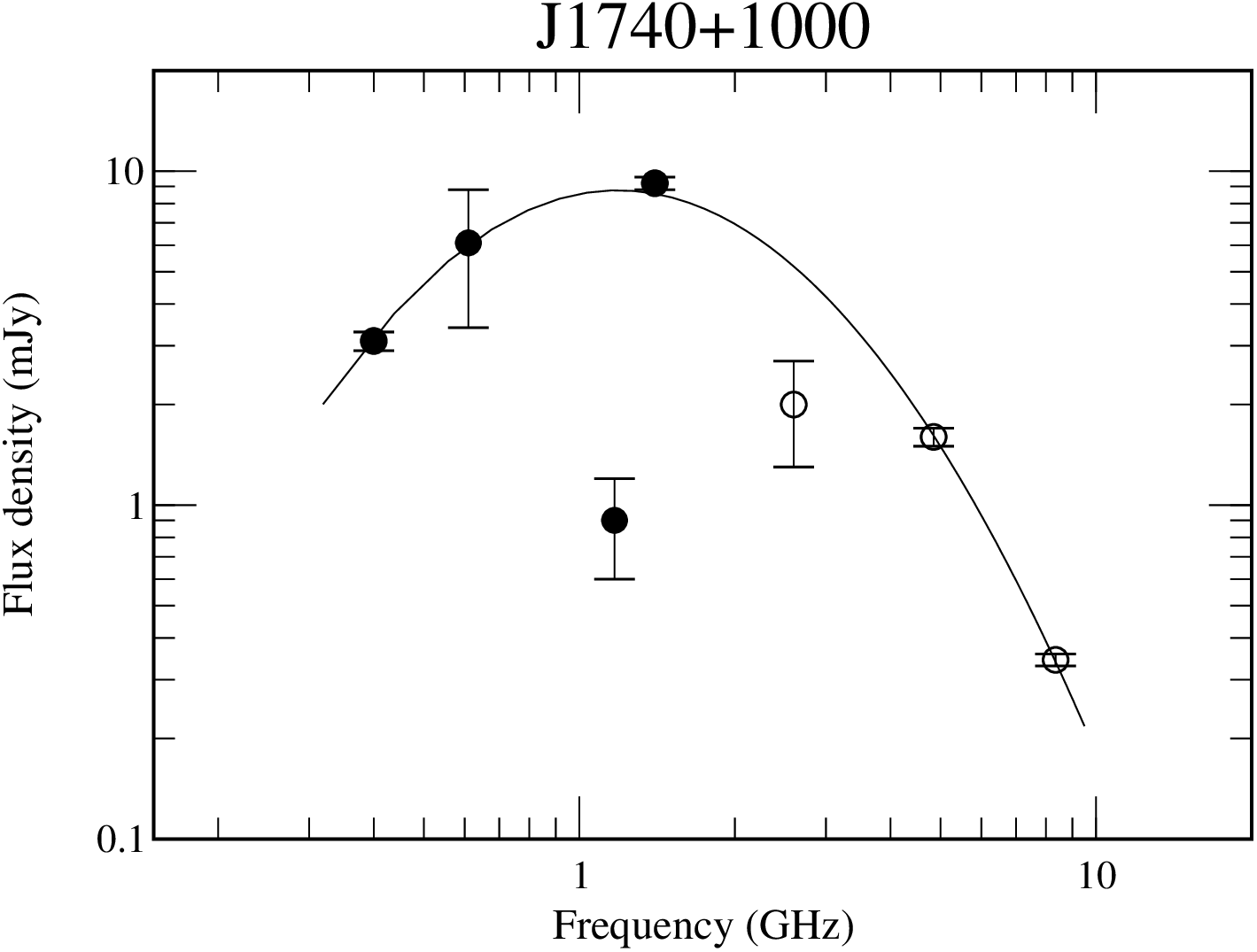}
\caption{The spectrum of PSR J1740+1000. Open circles denote our observations conducted 
in 2012, whereas data points marked with black dots are taken from~\citet{mcl2002} and 
\citet{kijak2011b}. The curve represents our fit to the data using the function given 
in Eq.~(\ref{turnover}). Flux density measured at 1070 MHz was excluded from the fitting 
procedure. Fitted parameters are: $a=$-$1.97\pm0.17$, $b=0.29\pm0.13$, $c=0.93\pm0.05$ 
with reduced $\chi^2=7.8$.}
\label{j1740}
\end{flushleft}
\end{figure}

\section{Spectra}
We constructed radio spectra for the pulsars listed in Table \ref{tab_flux}, using the 
flux density measurements derived from the above observing projects along with data taken 
from literature~\citep{lorimer95, kijak2007, kijak2011b, maron2000} and the ATNF catalogue. 
We divided our sources into groups depending on the morphological type of their spectra. 
The first two objects presented below can be classified as new GPS pulsars. Three other 
sources were revealed to be strong candidates for pulsars with high frequency turnover 
in their spectra, or possibly -- as in the case of PSRs B1811+40 and B1904+06 -- pulsars 
with broken spectra. A third group contains all the remaining sources which show power-law 
spectra (sometimes with a potential break or low-frequency turnover feature) together with the objects whose morphological types of spectra cannot be identified as they have their flux density measured at only 1 or 2 frequencies. 

\begin{figure}
\begin{flushleft}
\includegraphics[width=8.2cm]{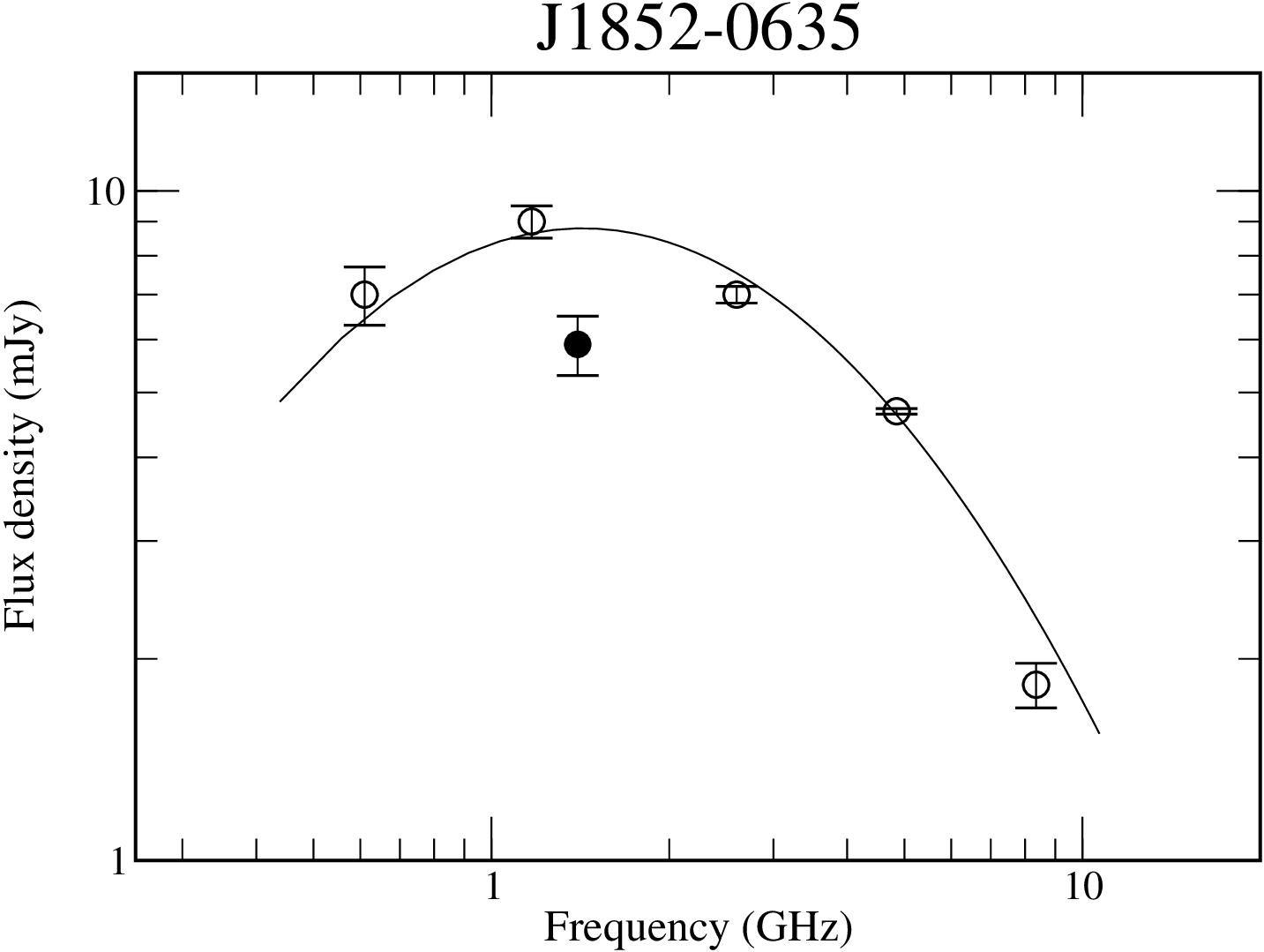}
\caption{The spectrum of PSR J1852$-$0635. Open circles denote results from our 
observations conducted in 2008 and 2010 (at 1070 MHz and 610 MHz, respectively; see also 
\citeauthor{kijak2011c}, \citeyear{kijak2011c}), and 2012 (at 2.6 GHz, 4.85 GHz and 8.35 
GHz), whereas the black dots denote data from the ATNF catalogue. The curve represents 
our fits to the data using the function presented in Eq.~(\ref{turnover}). Flux density 
measured at 1.4 GHz was excluded from the fitting procedure.  Fitted parameters are: 
$a=-0.99\pm 0.28$, $b=0.30\pm0.26$, $c=0.92\pm0.01 $, with reduced $\chi^2=10.8$.}
\label{j1852}
\end{flushleft}
\end{figure}

\begin{figure}
\includegraphics[width=8.2cm]{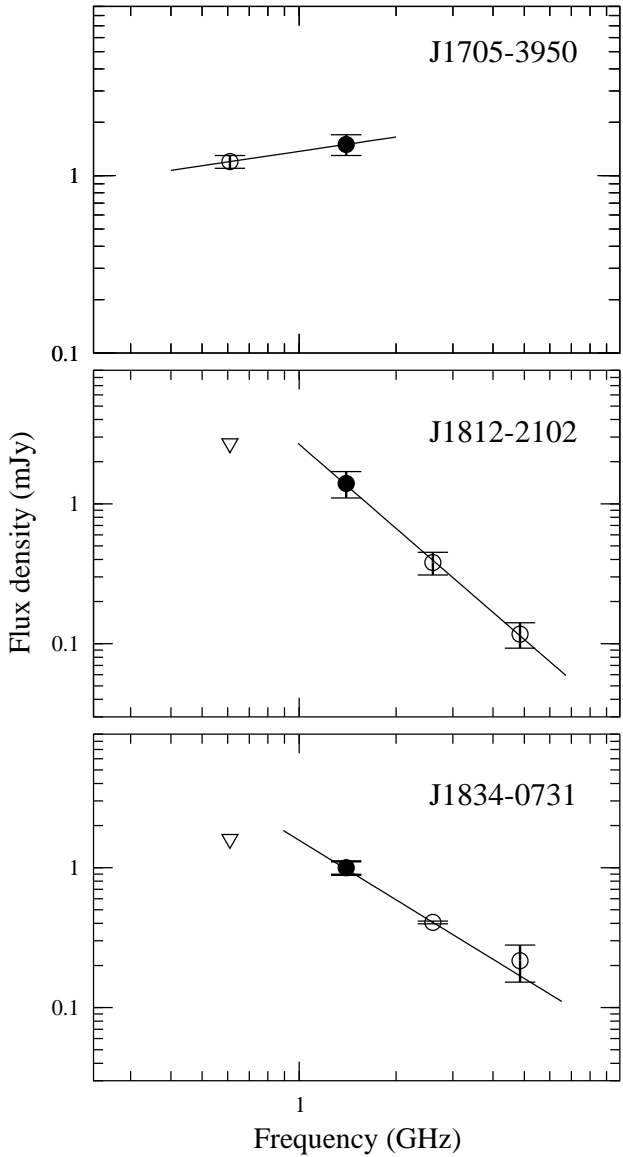}
\caption{The radio spectra of PSRs J1705$-$3950, J1812$-$2102 and PSR J1834$-$0731. Open 
symbols denote our observations conducted in 2010 (610 MHz) and 2012 (2.6 GHz and 4.85 GHz), 
whereas black dots denote data from the ATNF catalogue.  Triangles are upper limits. The 
straight lines represent our fits to the data using power-law function (upper limits were 
excluded from the fitting procedure).}
\label{candidates1}
\end{figure}

Graphical representation of these spectra 
are shown in Figures~\ref{j1740}-\ref{other2}. Open symbols denote new flux density 
measurements that resulted from our observing projects, whereas the previously published 
data are marked with black dots. Table~\ref{tab_flux} also shows the power-law spectral 
indices fitted to some of the presented spectra. The spectral fits to the data were 
carried out using an implementation of the nonlinear least-squares Levenberg-Marquardt 
algorithm.

\subsection{Newly-identified GPS pulsars}
\label{new_gps}
PSRs J1740+1000 and J1852$-$0635 have been identified as new pulsars with gigahertz-peaked 
spectra. In Fig.~\ref{j1740} and~\ref{j1852} we present their spectra along with our 
fits to the data using the function presented in Eq.~(\ref{turnover}).

\noindent\textbf{J1740+1000}\\
\citet{kijak2011b}  presented flux measurements of PSR J1740+1000 at frequencies of 610 MHz and 1170 MHz,  carried out with the GMRT in phased array mode. The authors were however unable to detect  the pulsar during their first attempt at high frequency observations using Effelsberg  telescope due to an erroneous ephemeris.  New measurements marked as open circles are  presented in Fig.~\ref{j1740}, along with previous results. One can note that the flux  density at 1170 MHz, derived from a previous observing project (see \citeauthor{kijak2011b}  \citeyear{kijak2011b}), appears to be substantially smaller than values obtained at  neighbouring frequencies -- this is possibly due to interstellar scintillations. Regardless of the reason causing the decrease of flux density at 1070 MHz, new observations appear to be sufficient to identify the PSR J1740+1000 spectrum as a GPS-type.\\

\noindent\textbf{J1852$-$0635}\\
\noindent PSR J1852$-$0635 which is shown to have positive spectral index at frequencies 
below 1 GHz~\citep{kijak2011c} seemed to be a strong candidate for GPS pulsar, which
is confirmed from our new results. Moreover, it is a relatively young object with high 
DM, as is typical of most of the GPS pulsars. To our knowledge there are no radio or 
high-frequency observations that would prove that this object has any kind of a peculiar 
environment, however one cannot rule out that possibility. 

In Fig.~\ref{j1852} we present the spectrum of PSR J1852+0635. It seems to be 
obvious that we are dealing with a GPS-type spectrum.  However, the object is a potential 
candidate for a pulsar with spectra described using the model presented in~\citet{l2008}. 
Thus, we attempted to fit three different functions that had been proposed in the literature, where
\begin{equation}
S(\nu)={S_0 \over 1+(2 \pi\nu\tau)^2}
\label{l_1}
\end{equation}
and
\begin{equation}
S(\nu)=S_0\left(1+(2 \pi\nu\tau)^2\right)^{n-1}\cdot e^{-i(n-1)\cdot atan(2\pi\nu\tau)}
\label{l_2}
\end{equation}
from \citet{l2008} are based on a nano-shot emission model. We also considered the 
empirical function presented in Eq.~(\ref{turnover}).
Parameters were
\begin{itemize}
\item $S_0=8.7 \pm 0.4 $ Jy, $\tau=0.03 \pm 0.003$~ns, $\chi^2=7.9$ for the function 
presented in Eq.~(\ref{l_1}),
\item $S_0=404 \pm 13 $ Jy, $\tau=0.038 \pm 0.005$~ns, $\chi^2=15.4$ for the function 
presented in Eq.~(\ref{l_2}),
\item $a=-0.99\pm 0.28$, $b=0.30\pm0.26$, $c=0.92\pm0.01 $, $\chi^2=10.8$ for the function 
presented in Eq.~(\ref{turnover}).
\end{itemize}
For (\ref{l_1}) we used a simplex fit procedure and a Levenberg-Marquardt method was 
found advantageous for (\ref{turnover}) and (\ref{l_2}).

The simple function presented in Eq.~(\ref{l_1}) provided the best fit with a nano-shot 
timescale of about 30~ps which is only 10\% of what was found for typical pulsars by
~\citet{l2008}. Alternatives (\ref{turnover}) and (\ref{l_2}) gave slightly worse 
agreements with the data, but generally speaking one finds that these three functions 
match the data quite well within the measured frequency interval. Only at frequencies 
above 20 GHz may one expect to be able to differentiate between them, with (\ref{l_1}) 
and (\ref{turnover}) being steeper than (\ref{l_2}). All fits predict an average flux 
density of about 0.5 - 0.8 mJy.

\subsection{Candidates for GPS pulsars}

Fig.~\ref{candidates1} presents the spectra of PSRs J1705$-$3950, J1812$-$2102 and 
J1834$-$0731, which can be considered good GPS candidates. New measurements points 
(open circles) along with the data from the ATNF catalogue (black dots) show negative 
spectral index at frequency ranges above 1 GHz for these objects. 

PSR J1705$-$3950, while having flux measurements at only two frequencies, seems to be 
a strong GPS candidate since it clearly shows a moderately high positive spectral index 
at frequencies close to 1 GHz. The spectra of PSRs J1755$-$2521 and J1857+0134 (not 
presented in Fig.~\ref{candidates1}) are similar to the spectrum of PSR J1705$-$3950, 
although in those cases we have only one flux measurement and one upper limit (see 
Table ~\ref{tab_flux}). Nevertheless, these two objects can be also considered good 
GPS candidates that deserve further investigation.

\begin{figure}
\includegraphics[width=8.1cm]{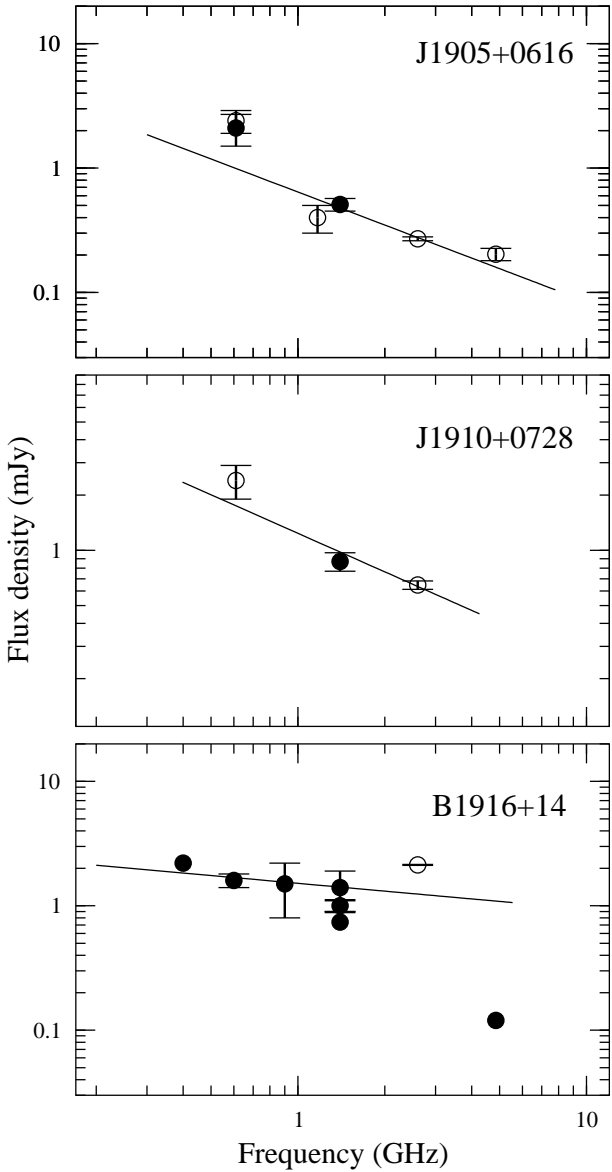}
\caption{The radio spectra of three pulsars: PSRs J1905+0616, J1910+0728 and B1916+14.  
Open circles denote our observations conducted in 2008 and 2010 (1070 MHz and 610 MHz, 
respectively), and 2012 (2.6 GHz and 4.85 GHz), whereas the black dots denote data 
from~\citet{maron2000}. The straight lines represent our fits to the data using a 
power-law function (in the case of PSR B1916+14 the marginal data point was excluded 
from the fitting procedure). The resulting spectral indices are presented in 
Table ~\ref{tab_flux}.}
\label{other1}
\end{figure}

In the cases of PSRs J1812$-$2102 and J1834$-$0731, from a consideration the level of 
the upper limit at 610 MHz and flux density measurements at higher frequencies (see Fig.~\ref{candidates1}), it appears that at frequencies below 1 GHz their spectra may be 
flat or even exhibit a positive spectral index.  Moreover, both objects are relatively 
young, and have high DMs which makes them similar to other GPS pulsars. However, it 
definitely requires further investigation to decide whether we are dealing with cases 
of turnover or broken type of spectra. The spectral indices fitted to high frequency 
flux values for these two pulsars (excluding data points that are upper limits) are 
presented in Table~\ref{tab_flux}: the values of $-$2.0 and $-$1.41 seem to indicate 
regular steep spectra; however, it was shown by Kijak et al. (2011) that, in the high
frequency range, most of the verified GPS pulsars behave like normal spectra pulsars.

\subsection{Other pulsars}
In this section we present the spectra of pulsars chosen from the remaining 16 objects
in our sample. For those with four and more observing points there is a strong suggestion 
that their spectra do not show any signs of the GPS. As for the spectra of 
the remaining objects it was impossible to properly classify them due to the deficiencies 
of the flux measurements. 

\begin{figure}
\includegraphics[width=8.4cm]{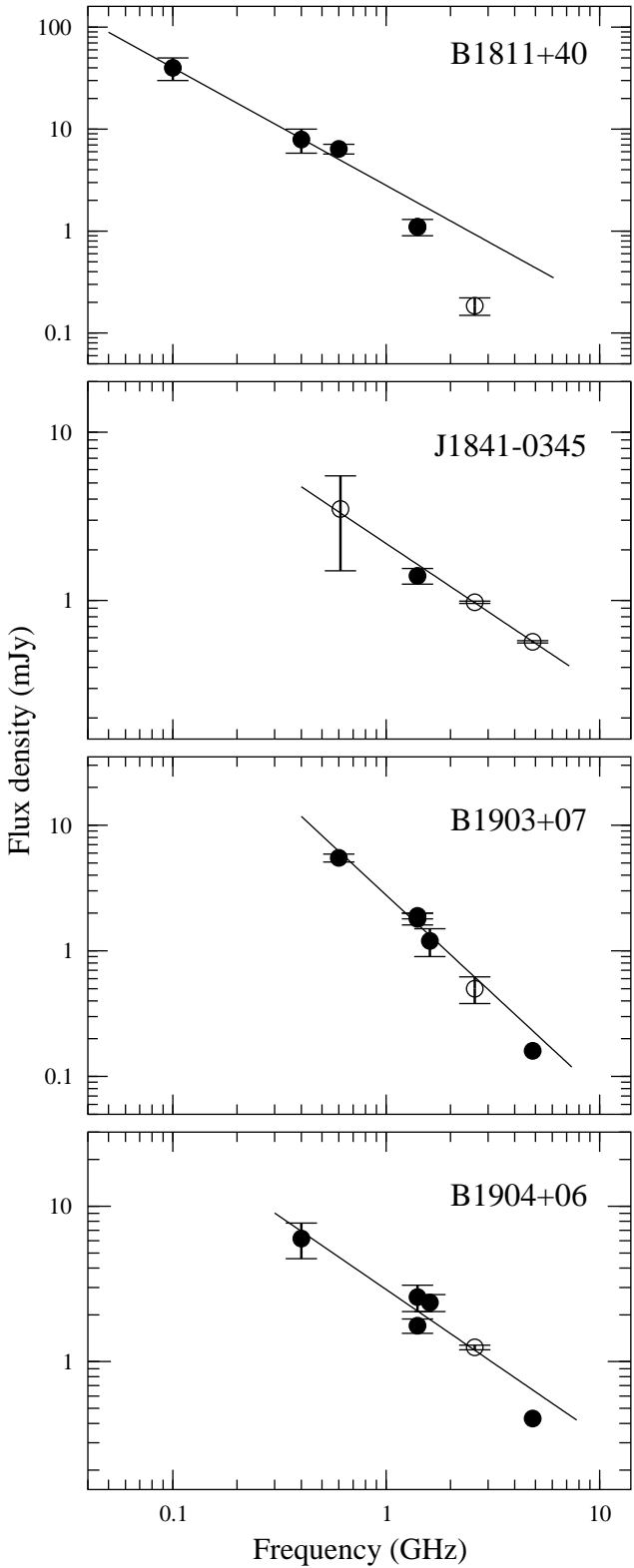}
\caption{The radio spectra of four pulsars: PSRs B1811+40, J1841$-$0345, B1903+07 and 
B1904+06.  Open circles denote our observations conducted in 2010 (610 MHz) and 2012 
(2.6 GHz and 4.85 GHz), whereas the black dots denote previously published data.  The 
straight lines represent our fits to the data using a power-law function. The resulting
spectral indices are presented in Table ~\ref{tab_flux}.}
\label{other2}
\end{figure}

The spectra for 7 of these remaining 16 pulsars from Table ~\ref{tab_flux} are presented 
in Fig.~\ref{other1} and Fig.~\ref{other2}.  For these objects we are probably dealing 
with a single power-law spectrum (the fits are presented along with data points). 
However, spectra of PSRs B1811+40 and B1904+06 presented in Fig.~\ref{other2} are 
potential candidates for broken spectra type with break frequencies 700 MHz and 2.3 GHz, 
respectively.

The data for PSR B1916+14, as in the case of PSR J1852$-$0635, was matched to the three 
different functions, and we obtained reasonable fits with the following parameters :
\begin{itemize}
\item $S_0=1.9 \pm 0.27 $ Jy, $\tau=0.1 \pm 0.03$~ns, $\chi^2=11.9$ for function 
presented in Eq.~(\ref{l_1}),
\item $S_0=202 \pm 303 $ Jy, $\tau=0.074 \pm 0.02$~ns, $\chi^2=13.5$ for function 
presented in Eq.~(\ref{l_2}),
\item $a=-1 \pm 0.3$, $b=-0.64 \pm 0.08$ $c=0.13 \pm 0.022$,  $\chi^2=10.7$ for 
function presented in Eq.~(\ref{turnover}).
\end{itemize}
Again, the fits only begin to diverge significantly above 20 GHz, but here average 
fluxes of the order of $\mathrm{\mu Jy}$ are to be expected, which will require a 
larger and more sensitive instrument that the currently available 100-metre class of 
radio telescopes.

In Table~\ref{tab_flux} we also include power-law spectral indices fitted for another
7 objects with spectra containing flux measurements at only two frequencies (PSRs 
J1723$-$3659, J1739$-$3023, J1744$-$3130, J1751$-$3323, J1835$-$1020, J1842$-$0905 and J1901+0510). For obvious reasons we did not conduct any fitting procedure for pulsars with flux density measured at only 1 frequency. 

One has to remember that almost all the above pulsars (with the only exceptions being 
PSRs B1811+40 and B1916+14) are objects with relatively high DMs. Therefore, using the 
standard pulsar flux measurement methods for these objects can cause erroneous results, 
especially when the observations are conducted at low observing frequencies where the 
pulsars are significantly affected by scattering. This typically happens when scattering 
tail is long enough to be a significant fraction of the pulsar period. In such cases 
one may overestimate the off-pulse level, which in turn would lead to an underestimate 
of the pulsar flux density.

\section{Summary and Discussion}

In this paper, we report two newly-identified GPS pulsars, PSR ~J1740+1000 and PSR ~J1852$-
$0635. We also present the spectra for 17 other pulsars, 13 of which are constructed for 
the first time. Among these, three sources can be considered as good candidates for 
gigahertz-peaked spectra pulsars.  Both the newly identified GPS pulsars are relatively 
young objects (115kyr for PSR~J1740+1000 and 567 kyr for PSR~J1852$-$0635), which  is 
a common characteristic for other known GPS pulsars. We have to note that PSR~J1740+1000 
is the first object exhibiting this phenomenon that has a low DM (24~pc~cm$^{-3}$). 

In case of PSR~J1740+1000 we present the spectrum over a wide frequency range, which is 
an improvement with regard to the previously published one \citep{kijak2011b}.  The new 
spectrum still contains the flux measurement at 1170~MHz (see Fig.~\ref{j1740}) which 
is unusually low compared to measurements at nearby frequencies, and we consider it to 
be unreliable and excluded it from the spectral fitting procedure. The reason for this
may be interstellar scintillations (as suggested in the discovery paper, \citeauthor{mcl2000} \citeyear{mcl2000}), 
and we plan to address this issue with future observations near the frequency of 1~GHz. 
For the other new GPS pulsar we only want to mention that the 1.4~GHz measurement (full 
circle in Fig.~\ref{j1852}) is also slightly lower than the shape of the spectra (based 
on our data) would suggest. This measurement was obtained by \citet{hobbs04} and 
published in the Parkes discovery paper.

For the majority  of the remaining pulsars we obtained spectral indices which range from 
+0.3 to $-$2.0. Six of these can be considered to indicate  flat spectra ($\xi>-1.0$). 
From our sample, we have selected three good GPS candidates. One of these sources is  
PSR~J1705$-$3950, with the spectral index $\xi=+0.3$ based on observations made at two 
frequencies only (see Fig.~\ref{other1}).  Because of its low declination we were unable 
to obtain flux measurements at high frequencies using the Effelsberg radio telescope; such 
measurements would be possible only using Australian telescopes. The other two candidates, 
PSR~J1812$-$2102 and J1834$-$0731, show typical spectra at high frequencies, but the 
flux density upper limits we obtained at  610~MHz suggest that the spectra deviate from 
a simple power-law which may be an indication of a possible GPS behaviour.

\subsection{GPS pulsars}

Summarizing, after the addition of the two new identifications presented here, we now 
have 11 GPS pulsars known. This number includes six objects presented in \citet{kijak2011a}, 
two radio magnetars \citep{kijak2013}, and PSR~ J2007+2722 presented by \citet{einstein}. 
This will definitely improve the statistics, which is important when trying to make 
reliable conclusions about general properties of this class of objects (both pulsars and 
radio magnetars).  However, the still relatively small sample of GPS pulsars, and no 
direct correlation between spectra shapes and physical parameters (internal or external) 
that could affect the radiation properties implies a still somewhat limited understanding
of the properties of these objects as a group.  Thus, successful searches for more 
GPS pulsars and low frequency flux measurements in radio pulsars are very important.

The first of the new GPS pulsars presented here, PSR J1740+1000, was discovered in an 
Arecibo survey ~\citep{mcl2000} and reported to have an unusually high positive spectral 
index of 0.9 ~\citep{mcl2002}, which made it a good candidate object. The new observations 
indeed appear to be sufficient to identify the PSR J1740+1000 spectrum as a GPS-type. 
The object is very interesting from several points of view. XMM-Newton observations have
revealed the existence of a PWN around the pulsar with a complex morphology (very 
extended pulsar tail, \citeauthor{k08} \citeyear{k08}). Later, \citet{k12} reported 
absorption features in the X-ray spectrum of this ordinary rotation-powered radio pulsar
and concluded that some of these features, thought to be unusual in nuetron star spectra,
are probably more common than earlier expected. 
The authors believe that their findings bridge the gap between the spectra of pulsars 
and other, more exotic neutron stars, i.e. ``X-ray dim Isolated NSs'' (XDINSs), Central 
Compact Objects (CCOs), and ``Rotating Radio Transients'' (RRATs).  PSR J1740+1000 was 
also considered by ~\citet{romani} as a candidate ``sub-luminous'' $\gamma$-ray pulsar. 
Additionally, PSR J1740+1000 reveals a new aspect of pulsars with gigahertz-peaked 
spectra: with a DM of 24 $\rm{pc\;cm^{-3}}$  (which along this line of sight corresponds 
to a distance of 1.2 kpc~\citeauthor{cordes2} \citeyear{cordes2}), it is much closer to us than other objects of this kind.

Due to the high dispersion measures of the GPS pulsars that were presented by
~\citet{kijak2011a} and~\citet{kijak2011b}, it has been suspected that high DM may have
an important role to play in producing the GPS feature in radio pulsar spectra. However,
with the recent GPS identifications reported here, it looks more likely that the effect 
may be related more to the environmental conditions in the close vicinity of these 
objects, rather than to the value of the DM. If that is the case, then the apparent 
observational trend for the GPS pulsars to prefer higher DM values may be a pure selection 
effect. As the GPS phenomenon is rather a rare occurrence it is obvious the chance for 
finding such objects increases with the size of the volume searched. This however does 
not exclude the possibility of finding such objects at relatively low distances. 

The case of PSR~B1259$-$63 shows that the GPS feature observed in the spectrum can be 
caused by the absorption effects in the near vicinity of the pulsar \citep
{kijak2011a,kijak2013}. Our results for  PSR ~J1740+1000 definitely support that idea, 
as it is another object with a peculiar environment, which makes it very similar to 
other GPS pulsars and magnetars. 

An alternative mechanism, the flux dilution (which is caused by anomalous scattering) 
may also lead to a decrease of energy  at lower frequencies, hence it may change the 
appearance of pulsar radio spectra (J. Cordes~--~private communication).  However, 
this would be somewhat difficult to explain the case of PSR~J1740+1000,  which is a 
relatively nearby source and hence likely to be less affected by strong scattering 
effects.

\section{Conclusions}

In their recent statistical study ~\citet{bates2013} concluded GPS sources may possibly 
constitute up to 10\% of the whole pulsar population, which could add up to as many as 
200 objects. We believe that the currently small size of the GPS pulsar sample does not 
come from the fact that they are a rare phenomenon, but rather from our limited knowledge 
of pulsar spectra in general, especially at frequencies below 1~GHz.

The GPS phenomena in radio magnetars along with the first low DM GPS pulsar 
reported here lead us to revisit our criteria of selecting candidate GPS sources. In 
the future searches, we would focus on sources with interesting (or extreme) environment 
rather than those with high DM.  This would make the objects with PWNs/SNRs found in 
high-energy observations excelent GPS pulsar candidates, and vice-versa, the sources 
with identified GPS would be good targets for high-energy observations.

Low frequency flux density measurements, which are crucial in searching for new GPS 
pulsars, can be difficult (or sometimes impossible) to conduct in the traditional way 
due to some of the known effects of interstellar medium.  Thus, we are able to determine 
pulsar flux densities using the standard on and off pulse energy estimates only when the 
scattering time is small enough. Many of the candidates for GPS pulsars have very high 
dispersion measures and at the lower frequencies they are not observable by means of 
regular pulsar observations, due to the extreme scattering which can smear their 
emission over the entire rotational period. This makes the observed radiation unpulsed 
(in extreme cases), or at least distorts the profile baseline to a degree that negates
the standard means of making the flux density measurement. For these cases, the only way 
to determine the pulsar flux is using interferometric continuum imaging techniques (see 
for example \citeauthor{kou2000} \citeyear{kou2000}). Thus, the interferometric imaging technique is important 
in these cases, providing an alternative for the typical pulsar observations and making 
possible verification of the gigahertz-peaked spectra, especially in the case of strong 
pulse scatter-broadening~\citep{lewandowski2013}.

\section*{Acknowledgments}
We thank the staff of the GMRT who have made our observations possible. The GMRT is 
run by the National Centre for Radio Astrophysics of the Tata Institute of Fundamental 
Research. This work is partially based on observations with the 100-m telescope of the MPIfR (Max-Planck-Institut 
f\"ur Radioastronomie) at Effelsberg.  This research was partially supported by the 
grants DEC-2012/05/B/ST9/03924  and DEC-2013/09/B/ST9/02177 of the Polish National 
Science Centre. MD was a scholar within Sub-measure 8.2.2 Regional Innovation Strategies, 
Measure 8.2 Transfer of knowledge, Priority VIII Regional human resources for the economy 
Human Capital Operational Programme co-financed by European Social Fund and state budget. 
The research leading to these results has received funding from the European Commission 
Seventh Framework Programme (FP/2007-2013) under grant agreement No 283393 (RadioNet3).

\section*{Appendix: Pulse profiles and widths}

In this appendix, we present pulse profiles for the observed pulsars collected during our observing projects and the width measurements of the profiles. 

\begin{figure}
   \includegraphics[width=8.2cm]{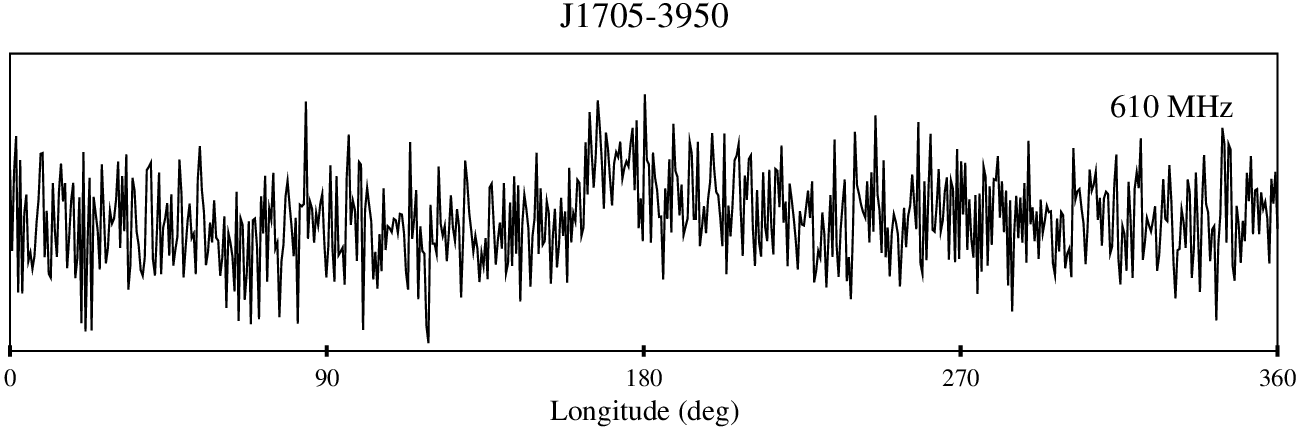}\vspace{0.45cm}
   \includegraphics[width=8.2cm]{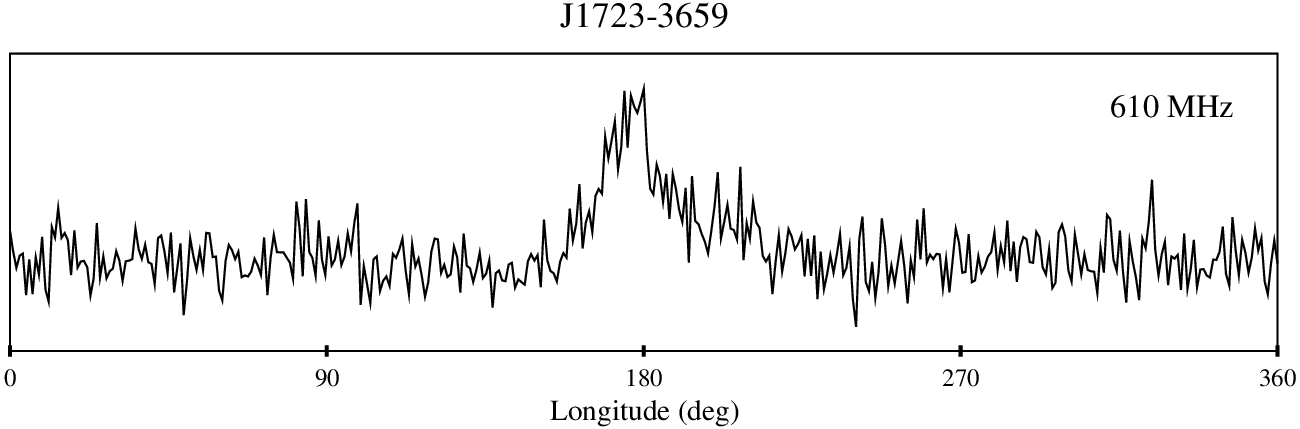}\vspace{0.45cm}
   \includegraphics[width=8.2cm]{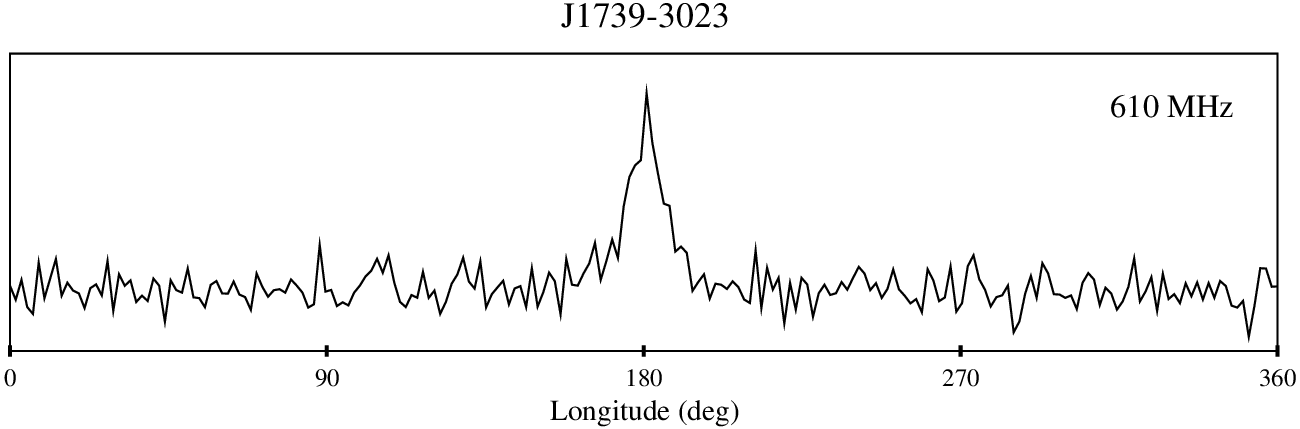}\vspace{0.45cm}
   \includegraphics[width=8.2cm]{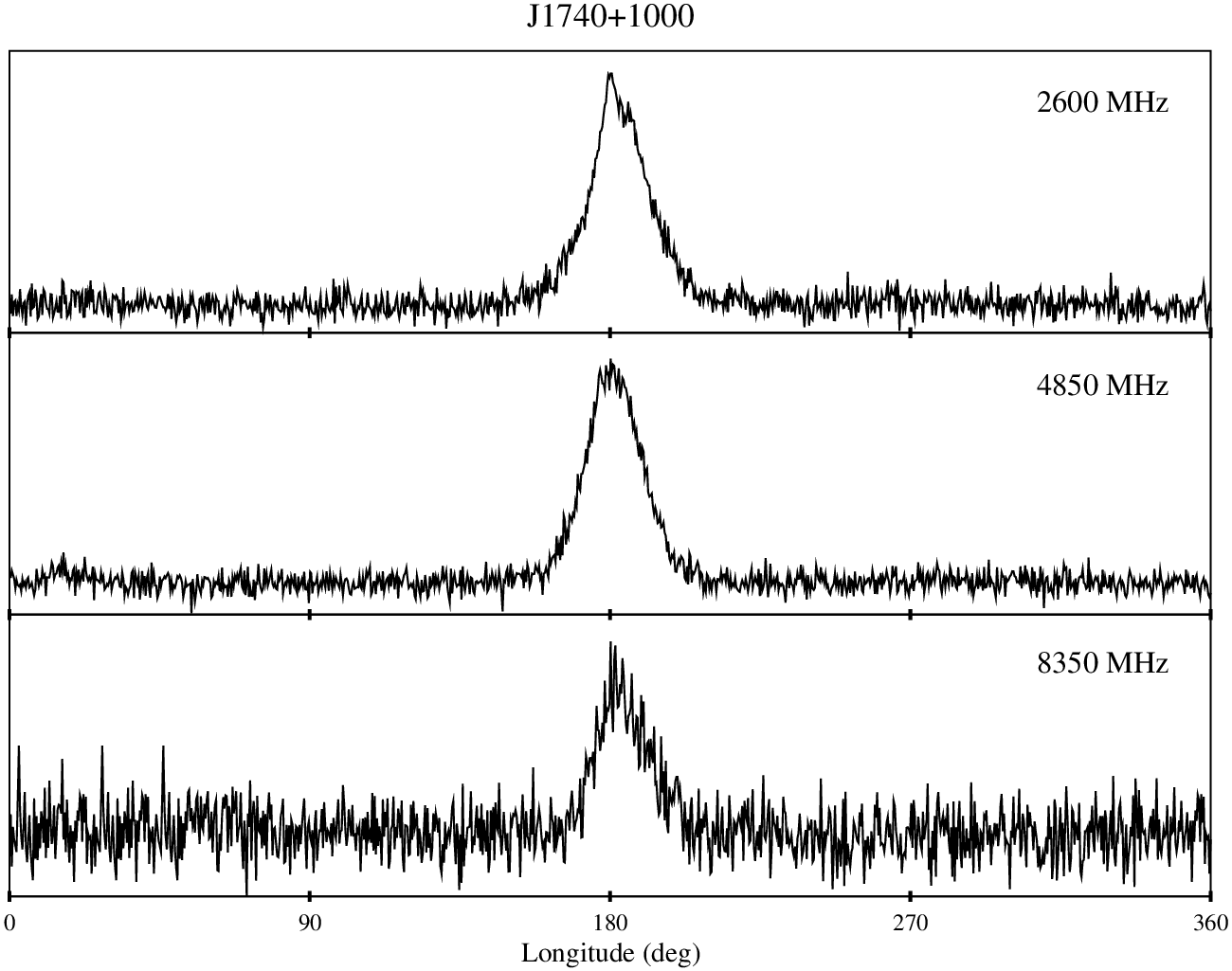}\vspace{0.45cm}
   \includegraphics[width=8.2cm]{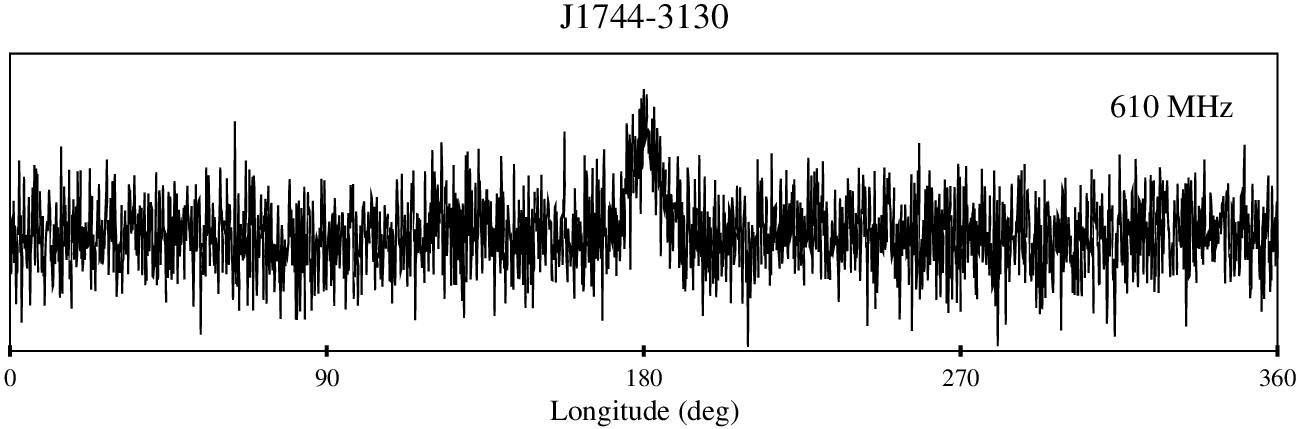}\vspace{0.45cm}
   \includegraphics[width=8.2cm]{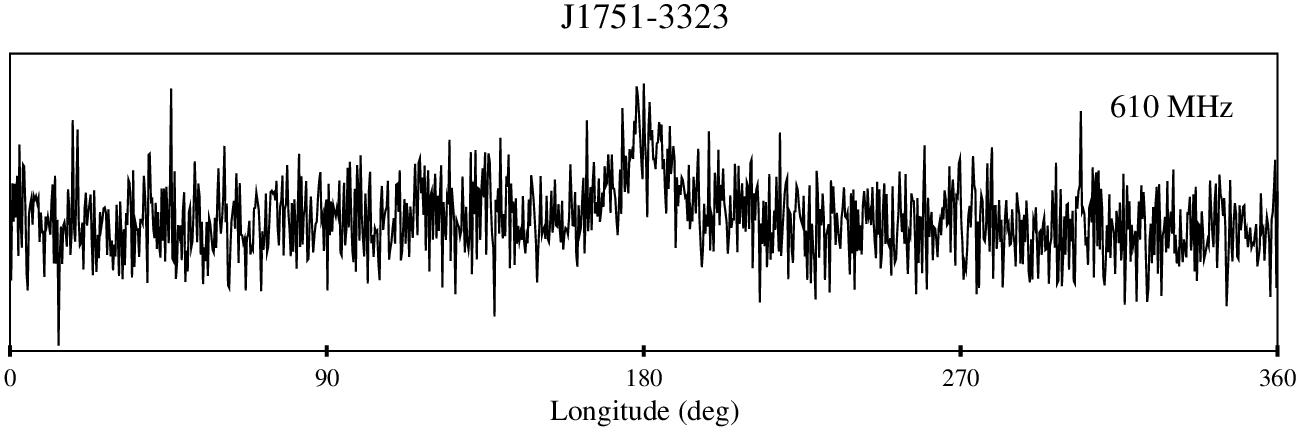}
   \caption{Pulsar profiles. Vertical axes present the flux density values in arbitrary units.}
   \label{profiles_1}
\end{figure}

\begin{figure}
   \includegraphics[width=8.2cm]{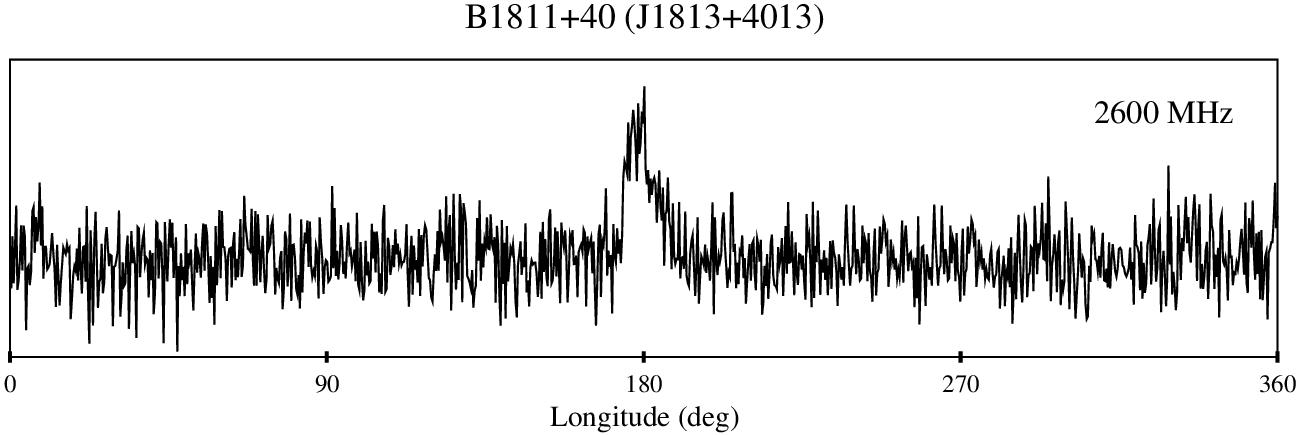}\vspace{0.45cm}
   \includegraphics[width=8.2cm]{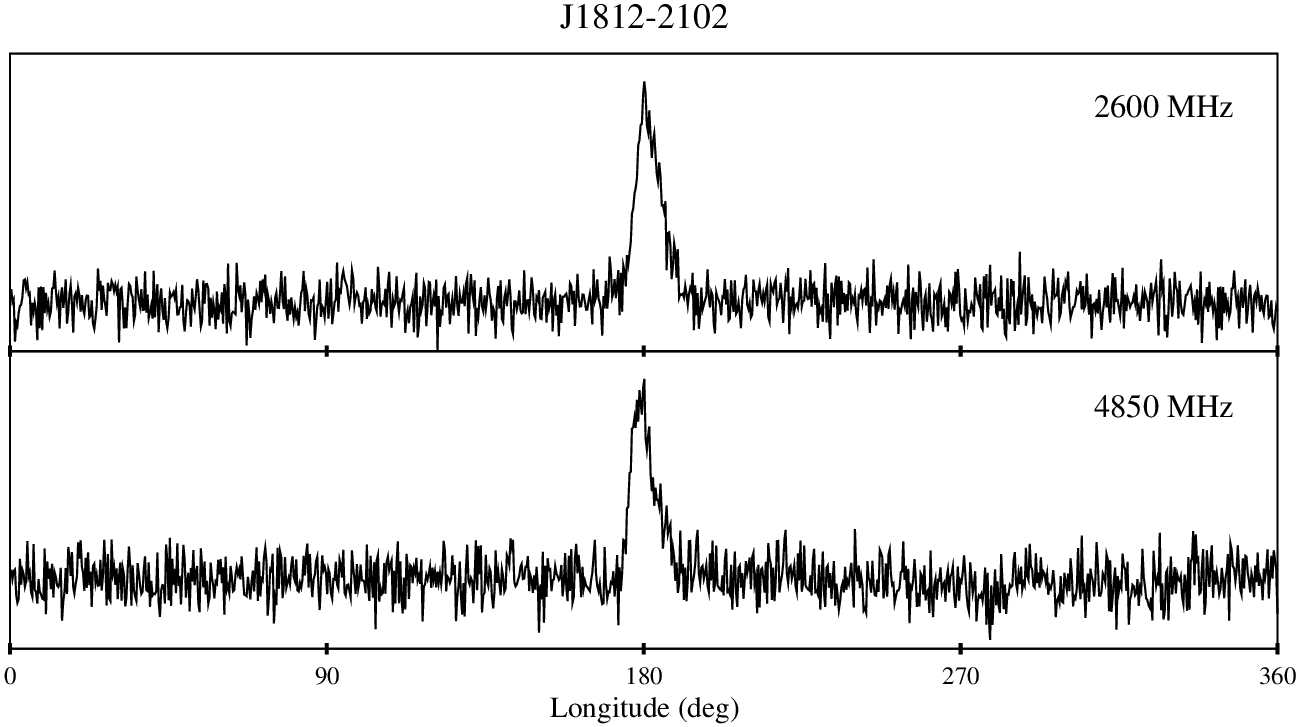}\vspace{0.45cm}
   \includegraphics[width=8.2cm]{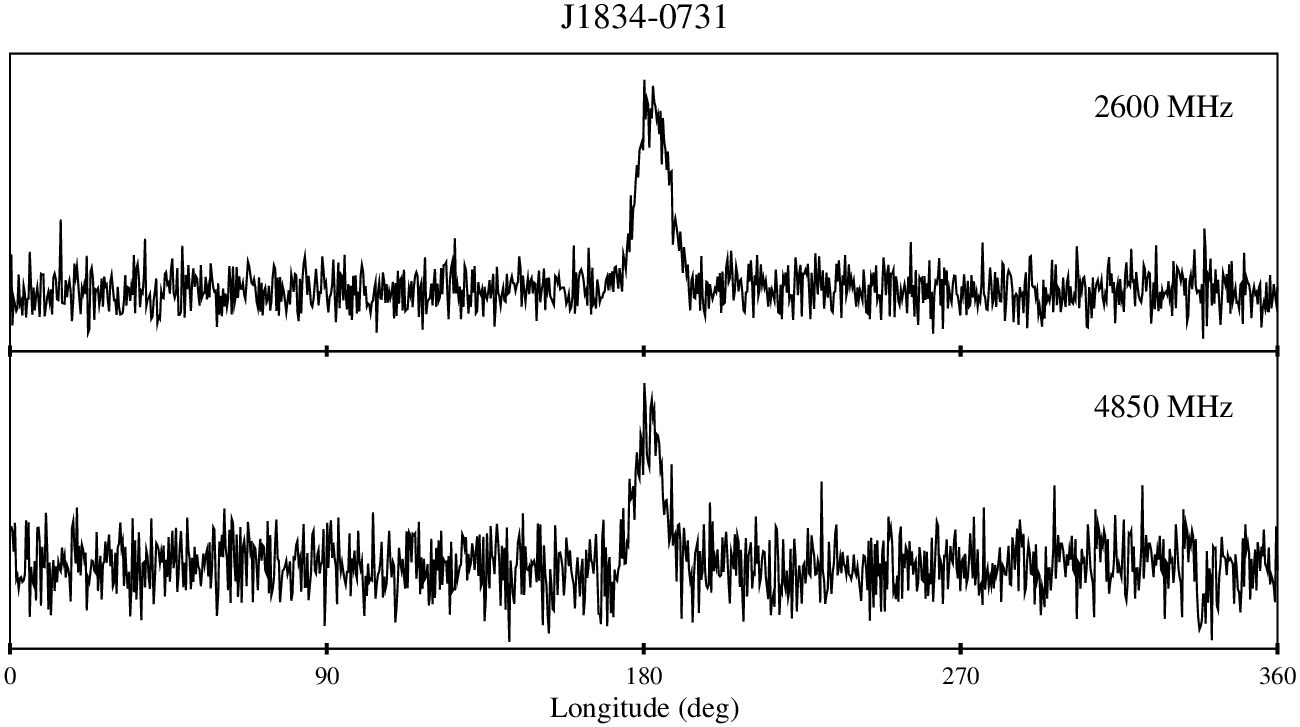}\vspace{0.45cm}
   \includegraphics[width=8.2cm]{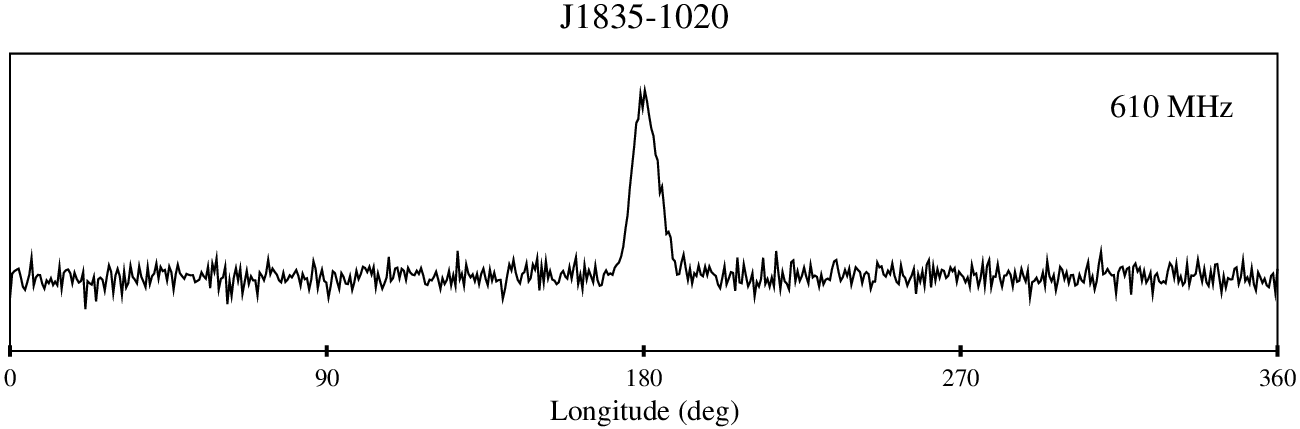}\vspace{0.45cm}
   \includegraphics[width=8.2cm]{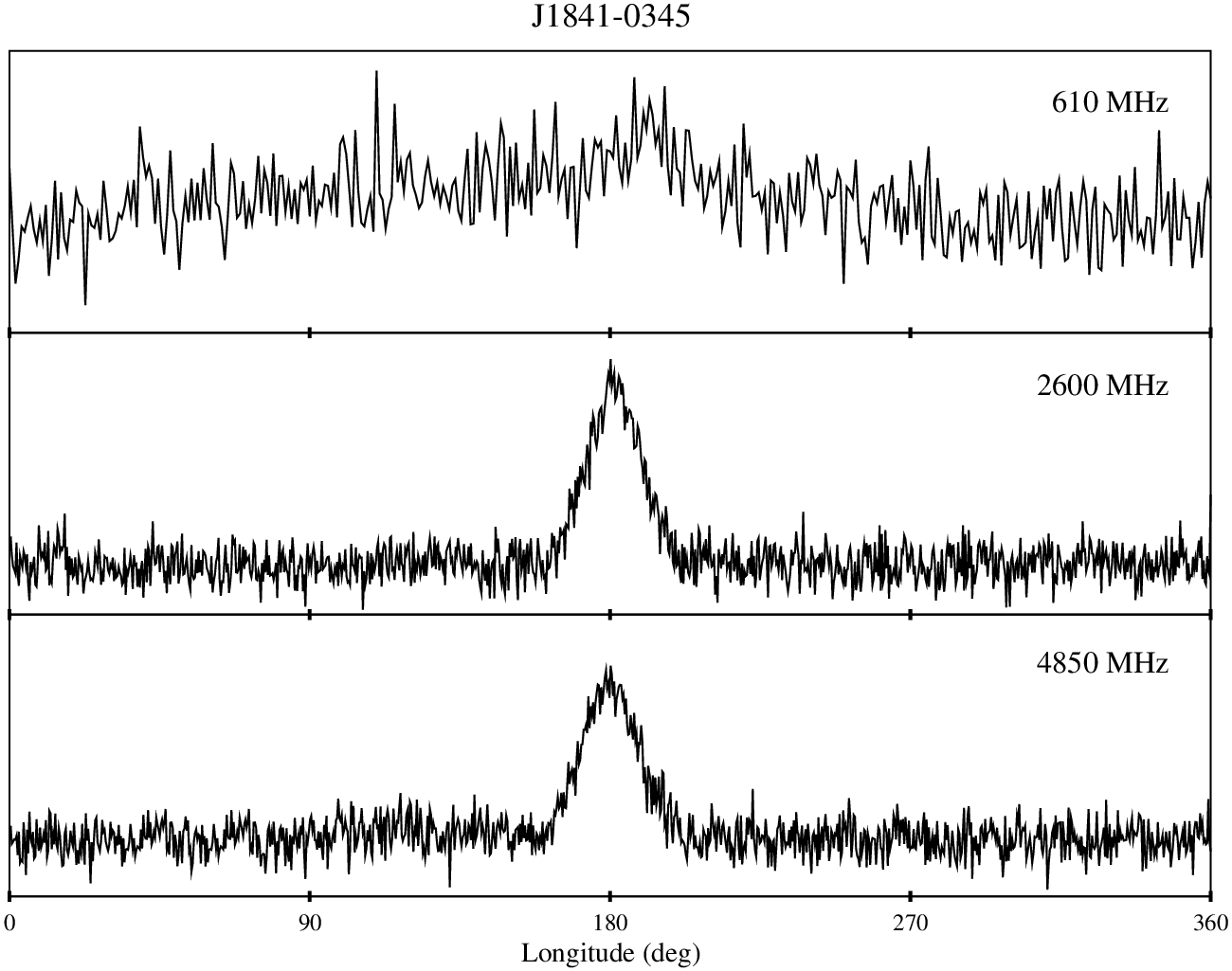}
   \caption{Pulsar profiles- continued. Vertical axes present the flux density values in arbitrary units.}
   \label{profiles_2}
\end{figure}

\begin{figure}
    \begin{flushleft}
   \includegraphics[width=8.2cm]{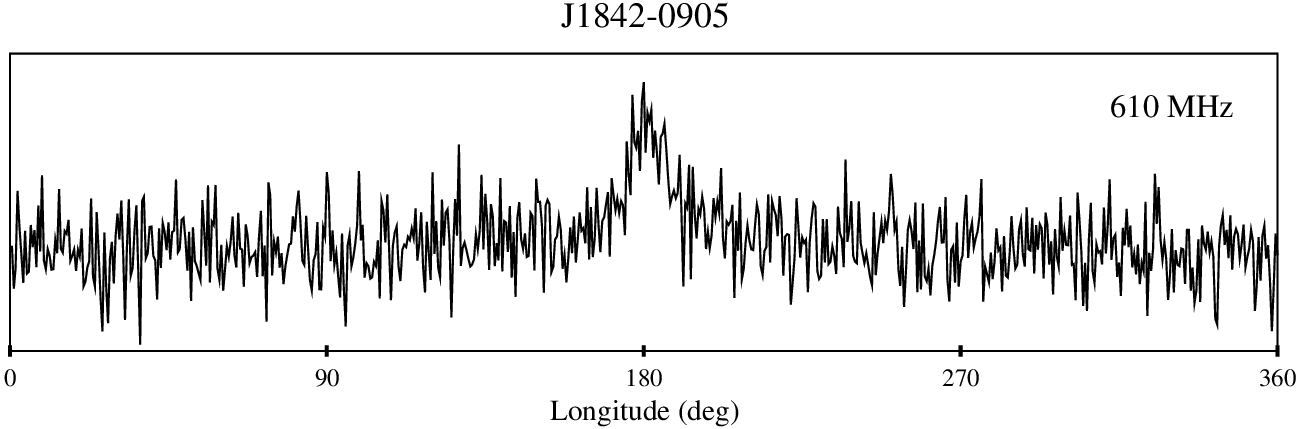}\vspace{0.2cm}
   \includegraphics[width=8.2cm]{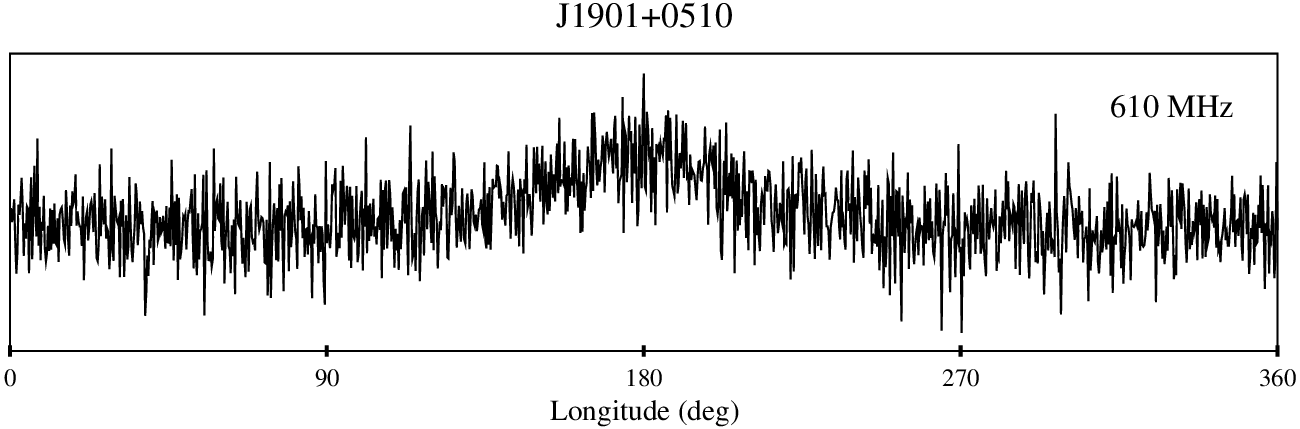}\vspace{0.2cm}
   \includegraphics[width=8.1cm]{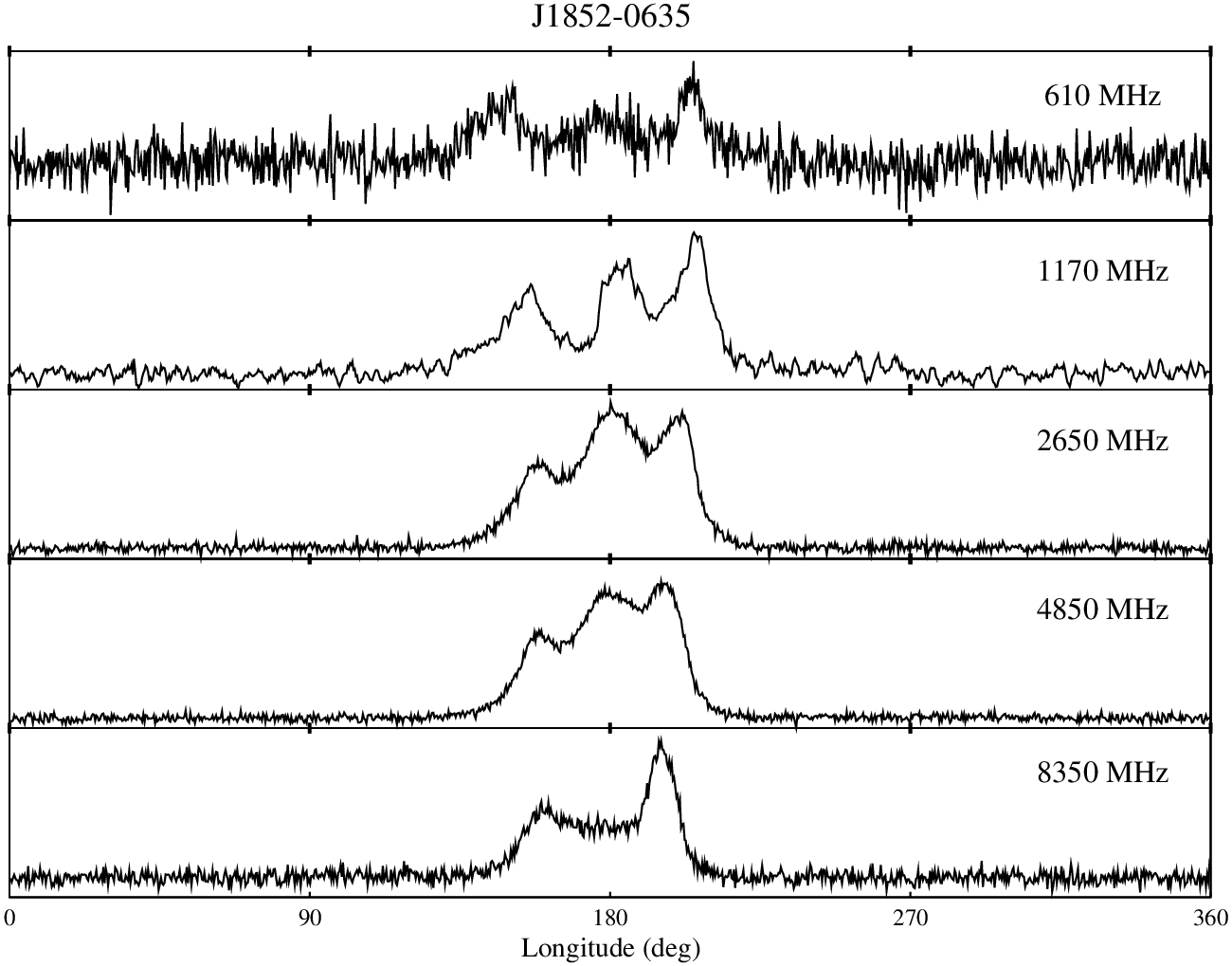}\vspace{0.2cm}
   \includegraphics[width=8.2cm]{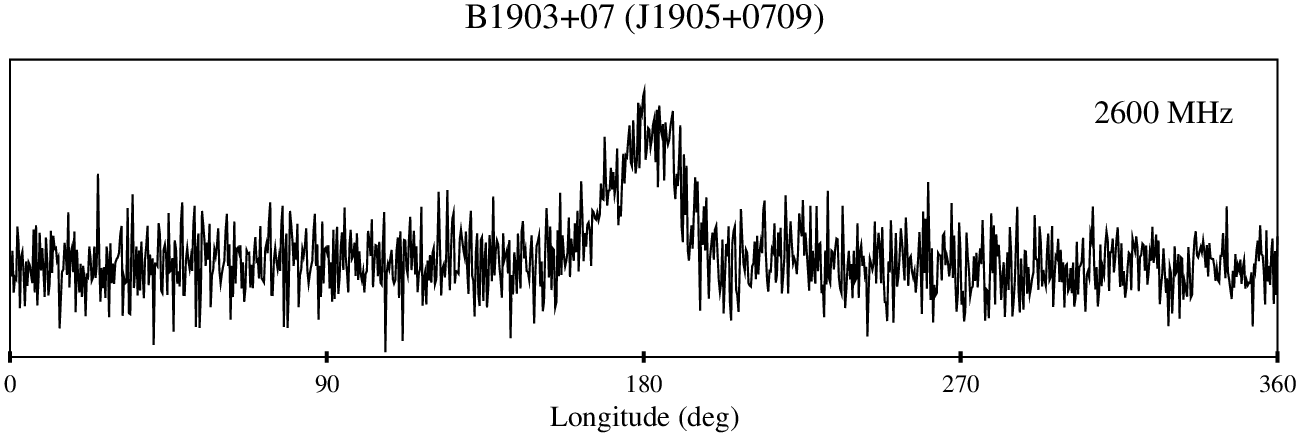}\vspace{0.2cm}
   \includegraphics[width=8.2cm]{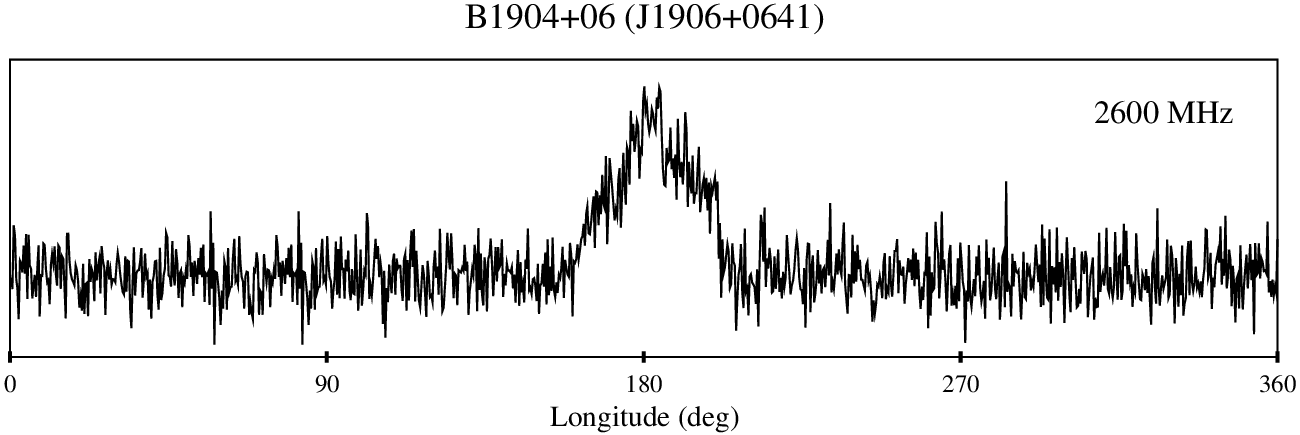}\vspace{0.2cm}
   \includegraphics[width=8.2cm]{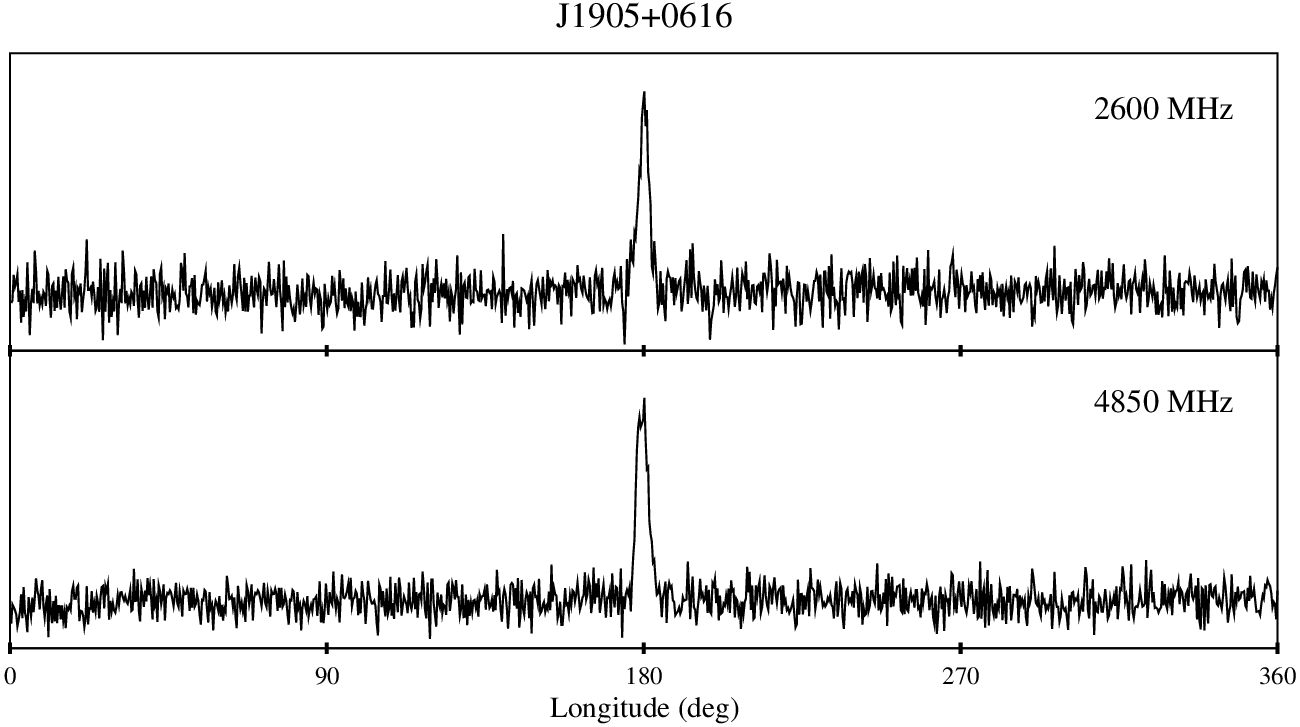}
   \caption{Pulsar profiles- continued. Vertical axes present the flux density values in arbitrary units.}
  \label{profiles_3}
   \end{flushleft}
\end{figure}

\begin{figure}
    \begin{flushleft}

   \includegraphics[width=8.2cm]{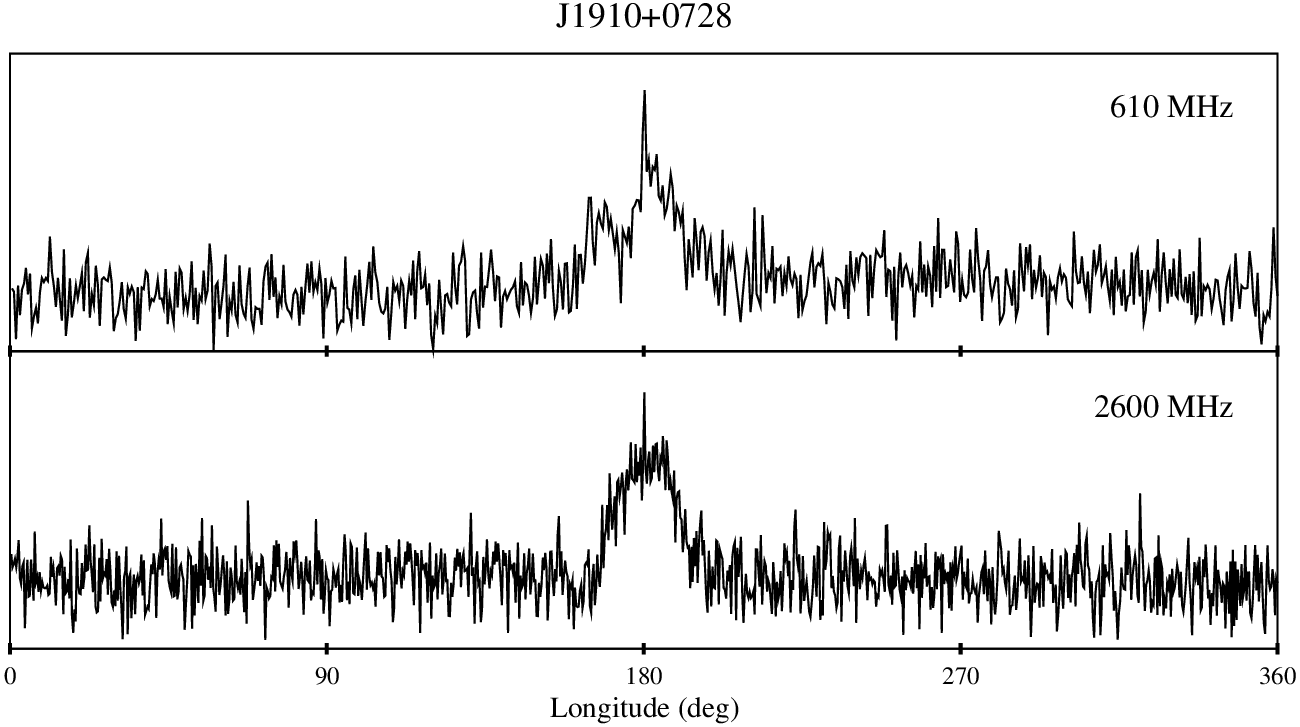}\vspace{0.4cm}
   \includegraphics[width=8.2cm]{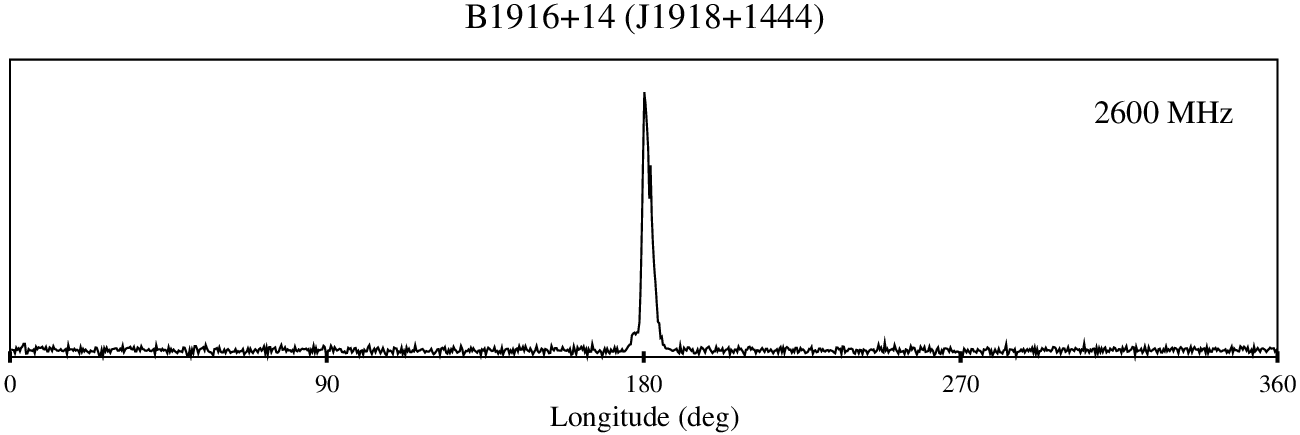}
   \caption{Pulsar profiles- continued. Vertical axes present the flux density values in arbitrary units.}
   \label{profiles_4}
   \end{flushleft}
\end{figure}

Figs.~\ref{profiles_1}-\ref{profiles_4} show the pulse profiles, acquired with both the GMRT (at 610 MHz and 1170 MHz) and the Effelsberg telescope (at 2600 MHz, 4850 MHz and 8350 MHz). We used the profiles to calculate the flux densities presented in Tab.~\ref{tab_flux}. The flux density values on the y-axes are expressed in arbitrary units.

The measurements of the pulse width are listed in Table~\ref{pulsars}.

\begin{table}
\caption{Pulse widths for observed pulsars.}
\centering
\begin{tabular}{l c D{.}{.}{-1} D{.}{.}{-1}}
\hline
Pulsar & Frequency & \multicolumn{1}{c}{$W_{10}$} & \multicolumn{1}{c}{$W_{50}$} \\
           & (MHz) & \multicolumn{1}{c}{(deg)} & \multicolumn{1}{c}{(deg)}\\
\hline
J1705$-$3905	& 610	& 69.9	& 38.3 \\
J1723$-$3659	& 610	& 42.2	& 23.1 \\
J1739$-$3023	& 610	& 23.8	& 13.1 \\
J1740+1000	& 2600	& 37.3	& 20.4 \\
			& 4850	& 36.3	& 19.3 \\
			& 8350	& 33.1	& 19.0 \\
J1744$-$3130	& 610	& 17.5	& 9.58 \\
J1751$-$3323	& 610	& 30.8	& 16.9 \\
B1811+40	& 2600	& 16.9	& 8.88 \\
J1812$-$2102	& 2600	& 15.5	& 8.49 \\
			& 4850	& 13.1	& 7.16 \\
J1834$-$0731	& 2600	& 17.2	& 9.42 \\
			& 4850	& 16.6	& 9.08 \\
J1835$-$1020	& 610	& 15.5	& 8.52 \\
J1841$-$0345	& 610	& 39.3	&21.5 \\
			& 2600	& 35.4	&19.4 \\
			& 4850	& 35.4	&19.4 \\
J1842$-$0905	& 610	& 25.2	&13.8 \\
J1852$-$0635	& 610	& 93.3	& 25.7\\
			& 1170	& 94.7	& 12.7 \\
			& 2600	& 70.4     & 37.4 \\
			& 4850	& 64.8     & 37.4 \\
			& 8350	& 59.6	& 11.5 \\
J1901+0510	& 610	& 111	& 61.0 \\
B1903+07	& 2600	& 37.4	& 20.5 \\
B1904+06	& 2600	& 49.4	& 27.1 \\
J1905+0616	& 2600	& 6.76	& 3.71 \\
			& 4850	& 6.77	& 3.72 \\
J1910+0728	& 610	& 46.1	& 25.3 \\
			& 2600	& 32.5	& 17.8 \\
B1916+14	& 2600	& 5.70	& 3.13 \\
\hline
 & & & \\
\end{tabular}
\label{pulsars}
\end{table}

\end{document}